\documentclass[a4paper,11pt]{article}
\pdfoutput=1 
\usepackage{jcappub}
\usepackage[T1]{fontenc}
\usepackage{appendix}
\usepackage{enumitem}
\usepackage{booktabs}
\usepackage{multirow}
\usepackage{siunitx}
\usepackage{orcidlink}

\def\l{\left}
\def\r{\right}

\title{\boldmath Neural networks and standard cosmography with newly calibrated high redshift GRB observations}

\author[a]{Celia Escamilla-Rivera\orcidlink{0000-0002-8929-250X},}
\author[b]{Maryi Carvajal,}
\author[a]{Cristian Zamora}
\author[c]{and Martin Hendry\orcidlink{0000-0001-8322-5405}}

\affiliation[a]{Instituto de Ciencias Nucleares, Universidad Nacional Aut\'onoma de M\'exico, Circuito Exterior C.U., A.P. 70-543, M\'exico D.F. 04510, M\'exico.}
\affiliation[b]{Universidad de Antioquia, Calle 70 No. 52-21. Apartado A\'ereo 1226. Antioquia, Colombia.}
\affiliation[c]{SUPA, School of Physics and Astronomy, University of Glasgow, Glasgow G12 8QQ, United Kingdom.}
\emailAdd{celia.escamilla@nucleares.unam.mx}
\emailAdd{maryi.carvajal@udea.edu.co}
\emailAdd{cristian.z.m@ciencias.unam.mx}
\emailAdd{Martin.Hendry@glasgow.ac.uk}

\abstract{
Gamma-ray bursts (GRBs) detected at high redshift can be used to trace the cosmic expansion history. However, the calibration of their luminosity distances is not an easy task in comparison to Type Ia Supernovae (SNeIa). To calibrate these data, correlations between their luminosity and other observed properties of GRBs need to be identified, and we must consider the validity of our assumptions about these correlations over their entire observed redshift range. In this work, we propose a new method to calibrate GRBs as cosmological distance indicators using SNeIa observations with a machine learning architecture. As well we include a new data GRB calibrated sample using extended cosmography in a redshift range above $z>3.6$.
An overview of this machine learning technique was developed in \cite{Escamilla-Rivera:2019hqt} to study the evolution of dark energy models at high redshift. 
The aim of the method developed in this work is to combine two networks: a Recurrent Neural Network (RNN) and a Bayesian Neural Network (BNN). Using this computational approach, denoted RNN+BNN, we extend the network's efficacy by adding the computation of covariance matrices to the Bayesian process. Once this is done, the SNeIa distance-redshift relation can be tested on the full GRB sample and therefore used to implement a cosmographic reconstruction of the distance-redshift relation in different regimes. Thus, our newly-trained neural network is used to constrain the parameters describing the kinematical state of the Universe via a cosmographic approach at high redshifts (up to $z\approx 10$), wherein we require a very minimal set of assumptions on the deep learning architecture itself that do not rely on dynamical equations for any specific theory of gravity.}
 
\begin{document}
\maketitle
\flushbottom


\section{Introduction}
\label{sec:intro}

The accelerated expansion of the Universe can be identified using the luminosity of Type Ia Supernovae (SNeIa), comprehensive measurements 
of which have been compiled over the last few years \cite{Barbary:2010bv,Betoule:2014frx}. However, due to their lack of sufficiently luminous observable at higher redshift, combined with constraints on the redshifts at which they form, most of these SNeIa are observed at $z\approx 1$ -- which further restricts our ability to distinguish between different cosmic expansion histories in a model-independent way. In this context, the SNeIa Pantheon compilation \cite{Scolnic:2017caz} is the only available sample that reports an observed SNIa at high redshift ($z=2.3$), with only six further SNeIa observed at $z\approx 1.4$. Hence, SNeIa will not be sensitive to differences between the luminosity distance redshift relation at higher redshifts in different cosmological models, and it is therefore important to identify cosmological probes that can be extended to higher redshifts. Furthermore, the robustness of the SNeIa calibration has also been questioned due the tension on the local value of the Hubble constant, $H_0$, with calls for a revision of the Cepheid period-luminosity relationships \cite{Breuval_2021} alongside a deep study of the latest Gaia Data Releases \cite{Lindegren_2021} and other parallax measurements from the Hubble Space Telescope (HST). 

On the other hand, Gamma-ray bursts (GRBs) are the most energetic cosmic explosions observed to date and are detectable up to very high redshift \cite{Cucchiara:2011pj}. Since their characteristics include a narrow range of  intrinsic luminosities, GRBs can be viewed as standard candles -- and hence calibrated to trace
the Hubble expansion at redshifts that extend well beyond those probed by SNeIa. However, calibrating the luminosity distances of GRBs is not a easy task due there being several different observables and physical parameters associated with their afterglow emission. Among the calibrating methods that have been employed in the literature are correlations between: the GRB spectrum lag and isotropic peak luminosity ($\tau_{\text{lag}}- L$ relation) \cite{Norris:1999ni}; the peak energy of the $\nu F_{\nu}$ spectrum and the isotropic equivalent energy ($E_p -E_\text{iso}$ relation) \cite{Amati:2000ad};  the time variability and isotropic peak luminosity ($V-L$ relation) \cite{Fenimore:2000vs}, the peak energy of the $\nu F_{\nu}$ spectrum and the isotropic peak luminosity ($E_{\rm p} -L$ relation) \cite{Amati:2009ts}; the minimum rise time of light curve and isotropic peak luminosity ($\tau_{\rm RT}- L$ relation) \cite{Schaefer:2006pa} and the peak energy and collimation-corrected
energy $(E_{\rm p} - E_\gamma)$ \cite{Ghirlanda:2005bz}. All of these correlations depend on a specific cosmological model that can lead to circularity issues when using the calibrated GRBs to constrain cosmological parameters. However, several model-independent methods have been proposed to avoid this issue, e.g. by calibrating the GRB relations first using a low-redshift subsample -- where there exists a corresponding subsample of SNeIa the luminosity distances of which can be fitted in a model-independent way -- and then extrapolating the correlations to high redshift.

Following these ideas, in this work we set out to use GRBs to understand the physics of the cosmic acceleration based on a model-independent description. Whichever path we take, we seek to avoid the explicit assumption of a cosmological model since this may lead to a biased reconstruction of the dynamics -- affecting directly our inference of the cosmological parameters as a result of  degeneracy problems or overfitting convergence using surveys. These problems are related to the \textit{cosmological tensions}, like those of $\sigma_8$ and $H_0$ \cite{DiValentino:2020zio,DiValentino:2020vhf}.

A reasonable strategy towards a model-independent panorama requires
only the postulation a priori
of homogeneity and isotropy, where the dynamics are independent of the energy densities. One path, then, is the so-called \textit{cosmography approach}, which has been shown to be a good method for reconstructing the kinematic evolution of the cosmic acceleration without assuming a cosmological model (see \cite{Capozziello:2013wha,Bolotin:2018xtq,Escamilla-Rivera:2019aol}, and references therein). To combine this approach and work with homogeneous data samples at low/high redshift, in \cite{Escamilla-Rivera:2019hqt} a new computational tool based on machine learning (ML) for supernovae, called Recurrent-Bayesian Neural Networks (RNN+BNN), was proposed. A deep learning architecture inside this ML can produce a trained homogeneous sample, e.g of luminosity distances.
Here, therefore, we extend that previous work and present a methodology to train supernovae and GRB data up to high-$z$ in order to test the viability of this cosmographic approach to reconstructing the distance-redshift relation. 
While this cosmographic approach has been previously studied from several points of view, setting limits on the standard cosmography using Taylor series \cite{Li:2019qic}, Pad\'e and Chebyshev polynomials \cite{Aviles:2012ay}, all of them have inherent convergence problems at higher redshifts -- and hence are vulnerable to bias and systematic errors. Also, this approach for late time observations has been considered in extended theories of gravity \cite{Capozziello:2019cav} and corrected versions of such theories as studied in \cite{Bahamonde:2021gfp}, where there are significant consequences for the reconstructions beyond the third order derivative in $H(z)$.

One interesting proposal to solve these issues was presented in 
 \cite{Munoz:2020gok}, where a generic EoS depending solely on the form of a dark energy-like term and its derivative was found. This approach can be adapted to any polynomial,
 to test the viability of cosmography since this result can be tested with a supernovae sample trained via a RNN+BNN network that solves problems with over-fitting at lower redshifts and increases the density of data in the $z$-region of interest. Furthermore, an as extension, in this work we design a new neural network to add a new GRB calibrated dataset, which can therefore help to compare the cosmographic $\mu(z)$ relation inferred from supernovae observations (assuming flatness) with the cosmographic $\mu(z)$ relation inferred from supernovae {\em and\/} GRBs, extending to higher forecast redshifts.
In the near future, high-redshift spectroscopic surveys \cite{Schlegel:2019eqc,EDGES} and/or gravitational-wave missions and projects \cite{Corman:2020pyr,LIGOScientific:2020stg} will provide accurate data in the redshift range of $2 < z <5$. Meanwhile, our trained RNN+BNN method will allow us to acquire a higher density of precise distance modulus data, for both SNeIa and GRBs, over the wide redshift range of $0.01 < z < 10$.
 
This paper is organised as follows: 
In Sec.\ref{sec:background} we present the theory behind the cosmographic approach in order to write a general expression for the luminosity distance in terms of the cosmographic parameters.
In Sec.\ref{sec:data} we describe the current cosmological observations: SNeIa and GRB samples, which will be used in the RNN+BNN methodology. Also we display the GRB values for the $\mu(z)$ function calibrated with the current Pantheon supernovae sample. 
In Sec.\ref{sec:RNN-BNN} we explain the basics behind our RNN+BNN method to train the cosmological data at hand, in order to the constrain the cosmographic parameters that appear in the definition of the luminosity distance. We consider an extension of the  Bayesian architecture to solve problems of e.g overfitting. Also, the method to include cosmography is detailed.
In Sec.\ref{sec:results} we present the results for the cosmographic analysis at high-$z$, and their behaviour according to different $H_0$ values. A Bayesian Information Criteria approach is then introduced to test the viability of the cosmography obtained.
Finally in Sec.\ref{sec:conclusions} we summarise and discuss our main results and conclusions.


\section{Cosmography background}
\label{sec:background}

According to the standard Cosmography paradigm, the only ingredient that we need to take into account \textit{a priori} in such an approach is a FLRW space-time derived from
kinematical requirements 
\begin{equation}
ds^2=-c^2dt^2+a^2(t)\l[\frac{dr^2}{1-kr^2}+r^2d\Omega^2\r],
\end{equation}
where $c$ is the speed of light, $a(t)$ is the scale factor and $k$ is a curvature constant. With this metric at hand it is possible to write the luminosity distance $d_L$
as a series expansion in the redshift parameter $z$, where the coefficients of the expansion are functions of $a(t)$ multiplying 
its higher order derivatives. With this expression, we can study the luminosity distance at high redshift in order to infer and constrain the cosmographic parameters
that best describe the model for our data.

At this point, the use of supernovae data (up to $z\approx 2$) is convenient; moreover the density of the data in luminosity distance can be enhanced if we consider also including GRBs. In particular, data samples coming from observations of both supernovae and GRBs can be used to fit directly a cosmographic expression for the luminosity distance $d_L$.

To construct this function, $d_L$ can be defined from the relation between the apparent luminosity, or flux, $l=L/4\pi d^{2}_{L}$ of an object, where $L$ is its absolute luminosity. In a FLRW space-time this can be described for small distances by the expression

\begin{eqnarray}
d_L(z)&=&\frac{c}{H_0}[z+\frac{1}{2}(1-q_0)z^2-\frac{1}{6}(1-q_0-3q_0^2+j_0+ \frac{kc^2}{H_0^2a^2(t_0)})z^3+ \nonumber \\ && +\frac{1}{24}[2-2q_0-15q_0^2-15q_0^3+5j_0+10q_0j_0+
s_0+\frac{2kc^2(1+3q_0)}{H_0^2a^2(t_0)}]z^4+\ldots]\,,
\end{eqnarray}
where the cosmographic parameters are defined as

\begin{eqnarray}
H_0 &\equiv&\frac{1}{a(t)} \frac{da(t)}{dt}|_{t=t_0} \equiv \frac{\dot{a}(t)}{a(t)}|_{t=t_0}~,\label{eq:H0}\\
q_0&\equiv&-\frac{1}{H^2}\frac{1}{a(t)}\frac{d^2a(t)}{dt^2}|_{t=t_0}\equiv-\frac{1}{H^2}\frac{\ddot{a}(t)}{a(t)}|_{t=t_0}~,\label{eq:q0}\\
j_0&\equiv& \frac{1}{H^3}\frac{1}{a(t)}\frac{d^3a(t)}{dt^3}|_{t=t_0}\equiv \frac{1}{H^3}\frac{a^{(3)}(t)}{a(t)}|_{t=t_0}~,\label{eq:j0}\\
s_0&\equiv& \frac{1}{H^4}\frac{1}{a(t)}\frac{d^4a(t)}{dt^4}|_{t=t_0}\equiv \frac{1}{H^4}\frac{a^{(4)}(t)}{a(t)}|_{t=t_0}~.\label{eq:s0}
\end{eqnarray}

Depending on the degree of the degree of Taylor series adopted, the cosmographic quantities derived may produce ambiguous or degenerate results. Moreover, this may lead to a misunderstanding in the definition of the proper distance -- although (as we discuss below) we may use the properties of the data themselves to limit the impact of this ambiguity through appropriate truncation of the Taylor series. In the rest of this work, we will refer to \textit{luminosity distance} as the most direct choice in the measure of distance for SNeIa and GRBs.

As a first attempt to fit the luminosity distance redshift relation using data samples, and to address the ill-conditioned behaviour at high-$z$ via appropriate convergence and truncation of the series, it is useful to recast $d_L$ as a function of the transformed variable $y=z/(1+z)$. In this manner, we can map $z\in(0,\infty)$ into the range $y\in(0,1)$, and retrieve an good
behaviour for the series at any distance (i.e. at any redshift). The consideration of this new variable $y$ does not change the definition of the cosmographic parameters but does change the equation expressing the luminosity distance as a series expansion in terms of $y$. The luminosity distance, at fourth order in $y$, can be written in terms of the cosmographic parameters as
\begin{eqnarray}
d_L(y)=&&\frac{c}{H_0}\left\{y-\frac{1}{2}(q_0-3)y^2+\frac{1}{6}\left[12-5q_0+3q^2_0-(j_0+\Omega_0)\right]y^3+\frac{1}{24}\left[60-7j_0-\right.\right.\nonumber
  \\ &&\left.\left.-10\Omega_0-32q_0+10q_0j_0+6q_0\Omega_0+21q^2_0-15q^3_0+s_0\right]y^4+\mathcal{O}(y^5)\right\}~,
  \label{eq:distance}
\end{eqnarray}
where $\Omega_0=1+kc^2/H_0^2a^2(t_0)$, is the total energy density. Using the latter expression, we can write the luminosity
distance \textit{logarithmic Hubble relation} as \cite{Lusso:2019akb}
\begin{eqnarray} 
\ln{\left[\frac{d_L(y)}{y~}\right]} \rm Mpc^{-1} =&&\frac{\ln{10}}{5}\left[\mu(y)-25\right]-\ln{y}=\ln{\left[\frac{c}{H_0}\right]}
-\frac{1}{2}(q_0-3)y+\frac{1}{24}\left[21-4(j_0+\Omega_0)+\right.\nonumber
  \\ &&\left.+q_0(9q_0-2)\right]y^2+\frac{1}{24}\left[15+4\Omega_0(q_0-1)+j_0(8q_0-1) 
  -5q_0+2q^2_0
  \right.\nonumber\\ &&\left.
  -10q^3_0+s_0\right]y^3+\mathcal{O}(y^4)~,\label{eq:y-exp}
\end{eqnarray}
and the corresponding expression for the distance modulus is given by
\begin{eqnarray}
\mu(y)=&&25+\frac{5}{\ln{10}}\left\{\ln{\left[\frac{c}{H_0}\right]}+\ln{y}
-\frac{1}{2}(q_0-3)y+\frac{1}{24}\left[21-4(j_0+\Omega_0)+q_0(9q_0-2)\right]y^2+\right.\nonumber
\\ &&\hspace{-.5cm}+\left.\frac{1}{24}\left[15+4\Omega_0(q_0-1)+j_0(8q_0-1)-5q_0+2q^2_0-10q^3_0+s_0\right]y^3+\mathcal{O}(y^4)\right\}~.
\label{eq:modulus}
\end{eqnarray}
The higher the order in the expansion of $d_L$, the better the fit to the data we can expect since there will be more free parameters -- but of course this comes at the cost of greater degeneracy in these parameters. However, for a given data sample there will be an upper bound on the order of the series which gives a statistically significant fit to those data.  With this bound, and our RNN+BNN approach, we will be able to objectively define a reasonable truncation to the expansion without having to simply cut the series by hand.

\section{The data}
\label{sec:data}

\subsection{Pantheon Type Ia Supernovae}
\label{sec:pantheon}

In this paper we select the recent Type Ia supernovae (SNeIa) compilation known as the  Pantheon sample \cite{Scolnic:2017caz}. This sample consists of 1048 SNeIa compressed in 40 redshift bins. It is the largest spectroscopic SNeIa sample compiled to date. 

As already noted, SNeIa can provide estimates of the distance modulus, $\mu$, the value of which is related to the luminosity distance $d_L$ as follows
\begin{equation}\label{eq:lum}
\mu(z)= 5\log{\left[\frac{d_L (z)}{1 \text{Mpc}}\right]} +25,
\end{equation}
where the luminosity distance is given in Mpc. In this distance modulus expression we should include the nuisance parameter, $\mathcal{M}$, as an unknown offset of the supernovae absolute magnitude (and including other possible systematics), which can also be degenerate with the value of $H_0$. As is standard, we assume spatial flatness and suppose that $d_L$ can be related to the comoving distance $D$ using
$d_{L} (z) =\frac{c}{H_0} (1+z)D(z),$
where $c$ is the speed of light. We obtain
\begin{equation}
D(z) =\frac{H_0}{c}(1+z)^{-1} \, 10^{0.2 \mu(z) -5}.
\end{equation}
The normalised Hubble function $H(z)/H_0$ is derived from the inverse of the derivative of $D(z)$ with respect to $z$, i.e.
\begin{equation}
D(z)=\int^{z}_{0} \frac{H_0 \, d\tilde{z}}{H(\tilde{z})}, \label{eq:dist}
\end{equation}
where $H_0$ is the value of the Hubble parameter we consider as a prior to normalise $D(z)$. For our sample, we calibrated the data by using 
a value obtained from late universe measurements.
corresponding to $H_0 =73.24 \pm 1.4 \, \, \mbox{km s}^{-1} \mbox{Mpc}^{-1}, $
from SH0ES + H0LiCOW,
with the corresponding value for the nuisance parameter
${\cal M}=-19.24 \pm 0.07$ \cite{Escamilla-Rivera:2019ulu}.

\subsection{Updated GRB dataset}

After the reconstruction calibration of $d_L$ from SNeIa, we can use them to calibrate the luminosity correlations
of GRBs. These correlations can be written by considering a general exponential form $R = AQ^b$, which can be
re-expressed in linear form as $y= a+ bx$, with $y\equiv \log R$, $x\equiv \log Q$ and $a= \log A$. Also, we can write the six luminosity correlations measured 
in the comoving frame as follows
\begin{eqnarray}
\log \left(\frac{E_\text{iso}}{\text{erg}}\right ) &=& a_1 + b_1 \log \left(\frac{E_{{\rm p},i}}{300\, \text{keV}}\right), \\
\log \left(\frac{L}{\text{erg}\,\text{s}^{-1}}\right ) &=& a_2 + b_2 \log \left(\frac{\tau_{{\rm RT},i}}{0.1\, \text{s}}\right), \\
\log \left(\frac{E_\gamma}{\text{erg}}\right ) &=& a_3 + b_3 \log \left(\frac{E_{{\rm p},i}}{300\, \text{keV}}\right), \\
\log \left(\frac{L}{\text{erg}\,\text{s}^{-1}}\right ) &=& a_4 + b_4 \log \left(\frac{E_{{\rm p},i}}{300\, \text{keV}}\right), \\
\log \left(\frac{L}{\text{erg}\,\text{s}^{-1}}\right ) &=& a_5 + b_5 \log \left(\frac{V_i}{0.02}\right), \\
\log \left(\frac{L}{\text{erg}\,\text{s}^{-1}}\right ) &=& a_6 + b_6  \log \left(\frac{\tau_{\rm lag},i}{0.1\, \text{s}}\right), 
\end{eqnarray}
with $V_i =V(1+z)$, $E_{{\rm p},i} = E_{\rm p} (1+z)$, $\tau_{{\rm lag},i} = \tau_{\rm lag} (1+z)^{-1}$ and $\tau_{{\rm RT},i} =\tau_{\rm RT}(1+z)^{-1}$, each of them obtained from the observations of GRB spectra.

To calibrate these six relations with our Pantheon sample, we consider that GRBs radiate isotropically by computing their bolometric peak flux $P_{\text{bolo}}=L/4\pi d^{2}_{L}$, where the uncertainty on
$L$ propagates from the uncertainties on $P_{\text{bolo}}$ and $d_L$. In the same manner, we can compute the isotropic equivalent energy $E_{\text{iso}}$ and the collimation-correlated energy $E_{\gamma}$, respectively as:
\begin{eqnarray}
E_{\text{iso}} = 4\pi d^{2}_{L} \, S_{\text{bolo}} (1+z)^{-1}, \\
E_{\gamma} \equiv E_{\text{iso}} (1-\cos \theta_{\text{jet}}),
\end{eqnarray}
where $\theta_{\text{jet}}$ is the jet opening angle and $S_{\text{bolo}}$ is the bolometric fluence of the GRBs. The GRB sample \cite{Wang:2015cya} used in this work consists of 116 long GRBs in the redshift range $0.17 \leq z \leq 8.2$.  With this sample we calibrated the luminosity correlations by maximizing the likelihood \cite{DAgostini:2005mth}
\begin{equation}
\mathcal{L}(\sigma_{\text{int}},a,b) \propto \prod_{i} \frac{1}{\sqrt{\sigma^{2}_{\text{int}}+\sigma^2_{yi} +b^2 \sigma_{xi}}} \times
\text{exp} \left[ -\frac{(y_i -a - bx_i)^2}{2(\sigma^2_{\text{int}} +\sigma^2_{yi} +b^2 \sigma^2_{xi})} \right].
\end{equation}
The best-fitting parameters and their uncertainties,$(a , b, \sigma_{\text{int}})$,  obtained in this way are reported in Table \ref{tab:grb}. With these fits we tabulated our full GRB sample, as reported in Tables \ref{tab:grb2} and \ref{tab:grb2-1}.

\begin{table*}
\centering
\begin{tabular}{|l|c|c|c|c|} 
 \hline 
Correlation 		         & $a$ 				& $b$ 				& $\sigma_{\text{int}}$ \\ \hline \hline
$E_{p}-E_{\text{iso}}$  &  $52.778\pm0.058$    & $1.546\pm 0.129$   & $0.461\pm 0.051$ \\
$\tau_{RT}-L$  &  $52.766\pm0.078$    & $-1.250\pm 0.132$   & $0.420\pm 0.055$ \\
$E_{p}-E_{\gamma}$  &  $51.655\pm0.065$    & $1.455\pm 0.156$   & $0.136\pm 0.010$ \\
$E_{p}-L$  &  $52.167\pm0.059$    & $1.450\pm 0.138$   & $0.531\pm 0.049$ \\
$V-L$  &  $51.785\pm0.140$    & $0.556\pm 0.131$   & $0.747\pm 0.069$ \\
$\tau_{lag}-L$  &  $52.366\pm0.071$    & $-0.781\pm 0.116$   & $0.452\pm 0.058$ \\
\hline 
 \end{tabular} 
 \caption{Best-fit values, and their uncertainties, for the parameters describing the linear correlations between the GRB observational data and intrinsic properties. In each case we performed an MCMC analysis to calculate the posterior probability density function of the parameter space with a flat prior on all free parameters and a fixed constraint of $\sigma_{\text{int}}>0$.}
 \label{tab:grb} 
\end{table*}

\begin{table*}
\centering
\begin{tabular}{|l|c|c|c|c|c|c|c|c|c|} 
 \hline 
$z$ 		         & $\mu$ 	   & $\sigma_{\mu}$  		&$z$  & $\mu$ 	   & $\sigma_{\mu}$    &$z$  & $\mu$ 	   & $\sigma_{\mu}$  
\\ \hline 
0.03351 & 35.1195 & 0.933638 & 1.44 &43.2858 &0.95932&2.199 &46.4216 &1.06071\\
0.125 & 38.6856 &1.84671&1.4436 &43.8132 &1.01313&2.2 &44.9272 &0.931098\\
0.17& 38.4667 &1.03564&1.46 &44.1083 &1.11878&2.20&45 46.8579 &0.931104\\
0.25& 39.9357 &1.38443& 1.48 &43.9951 &1.0662&2.22 &45.2389 &1.20041\\
0.3399 &40.6749 &0.95616& 1.48 &43.5582 &0.929143&2.26 &43.8782 &0.929903\\
0.36 &42.9717 &1.07033&1.489& 45.0134& 1.043&2.27 &45.029 &0.935913\\
0.41 &40.8059 &1.1032&1.52 &42.871 &0.978916&2.296 &45.502 &1.15448\\
0.414 &42.8313 &1.21619& 1.547 &45.9243 &0.946461&2.3 &46.2112 &1.2461\\
0.434 &41.5047 &0.985045& 1.547 &44.0644 &0.973486&2.346 &46.8641 &1.16259\\
0.45 &41.8452 &1.0036& 1.563 &42.781& 1.69995&2.433 &46.867 &1.09682\\
0.49 &39.7308 &0.946699&1.567 &43.8473 &0.927328& 2.452 &47.364 &1.13965\\
0.5295& 42.6995 &1.03825&1.6 &44.1755 &1.0596&2.486 &44.8287 &0.944518\\
0.54 &39.9036 &0.925708&1.604 &46.5316 &0.962389& 2.488 &45.181 &0.942494\\
0.542 &41.4047 &0.975989&1.608& 46.8731 &1.03718& 2.512 &45.5879 &0.934876\\
0.543 &42.2564 &1.32376&1.613 &44.6927 &0.948442&2.58 &44.4939 &0.943808\\
0.544 &42.0376 &0.927353&1.619 &44.3532 &0.946002&2.591 &46.1254 &0.958853\\
0.55 &42.5063 &0.951189&1.64 &45.0375 &0.958753&2.615 &45.9102 &1.00661\\
0.606 &41.6888 &0.940851&1.6919& 43.2837 &1.03666&2.65 &45.6525 &1.00258\\
0.618 &43.7538 &1.07668&1.71 &45.675 &0.960875&2.671 &45.3823 &0.943181\\
0.6528 &41.8718 &1.00894&1.727 &43.2599 &0.973643&2.69 &46.1013 &1.16024\\
0.677 &43.7533 &0.983363&1.728 &44.3889 &0.946023&2.752 &45.5087 &1.07352\\
0.689 &43.868 &0.93225&1.728 &44.3889 &0.946023&2.77 &46.0964 &0.9302\\
0.69 &42.2854 &0.928094& 1.77 &42.9739 &0.944986&2.821 &46.6406 &0.937652\\
0.695& 43.1984 &1.11582& 1.798 &44.5549 &0.94189&2.83 &46.6737 &0.930181\\
0.706& 41.2777 &0.945272&1.8 &45.3824 &0.948821&2.893 &47.598 &1.08191\\
0.716 &40.4242 &0.95484&1.822 &44.7661 &0.925467&2.8983 &45.3038 &1.02857\\
0.716 &41.2868 &1.01881&1.858 &45.1695 &1.23345&3. &46.1498 &1.10789\\
0.736 &44.393 &0.9305&1.858 &45.1695 &1.23345&3.036 &46.0503 &0.959184\\
0.78 &40.8528 &0.974974& 1.8836 &46.2739 &1.28842&3.038 &44.8963 &1.19331\\
0.8 &42.3122 &1.13123& 1.9 &45.7854 &1.13272&3.075 &46.7497 &1.03846\\
0.8049& 44.9536 &0.94066&1.9229& 44.6651 &1.11463&3.2 &45.7872 &1.11734\\
0.82 &42.816 &0.942884&1.95 &46.5089 &1.08858&3.22 &45.5887 &1.00027\\
0.835 &44.1715 &1.08963&1.9685& 44.6084 &0.999689&3.35 &47.5973 &0.954116\\
0.842 &42.7009 &1.09192&1.98 &44.5553 &1.02204&3.36 &47.7168 &1.01118\\
0.846 &43.8624 &0.92956&2.05 &46.8257 &1.31834&3.37 &47.4274 &1.22111\\
0.859 &43.1051& 1.13207&2.07 &43.9293& 0.936188&3.42 &47.0429 &0.996013\\
0.8969 &44.5543 &0.999868&2.0858 &44.6893 &1.56663&3.424 &47.0792 &1.05247\\
0.937 &42.9071 &0.931116&2.088 &45.8633 &0.983381&3.5 &46.7831 &1.23155\\
0.947 &42.1526 &1.03421&2.106 &46.3176 &0.965919&3.5 &45.2884 &0.974999\\
0.958 &42.9866 &1.00157&2.1062 &43.7433 &0.925125&3.5328 &46.978 &1.1415\\
0.966 &44.2503 &0.957411&2.14 &44.7932 &0.957566&3.57 &45.8457 &0.986639\\
0.971 &42.7535 &1.19374 &2.145 &47.4073 &0.984835&3.6 &45.6286 &1.09246\\
\hline  
 \end{tabular} 
 \caption{Compilation of estimated GRB luminosity distance moduli, and their errors, for redshifts $z<3.6$}.
 \label{tab:grb2} 
\end{table*}

\begin{table*}
\centering
\begin{tabular}{|l|c|c|c|c|} 
 \hline 
$z$ 		         & $\mu$ 		& $\sigma_{\mu}$ \\ \hline 
3.758 &47.5053 &0.963238\\
3.796 &46.1613 &1.12782\\
3.91 &46.6447 &1.11229\\
3.93 &47.0881 &1.17098\\
4.0559 &48.3284 &1.27013\\
4.109 &46.899 &1.19876\\
4.1745 &47.4784 &1.00188\\
4.35 &47.145 &0.960048\\
4.5 &46.1465 &1.22279\\
4.6497 &49.1939 &1.1254\\
5.46 &47.1793 &1.06039\\
6.295 &49.5673 &1.13213\\
6.695 &49.788 &1.31553\\
8.1 &49.2684 &1.0345\\
9.3 &50.0158 &0.959439\\
\hline  
 \end{tabular} 
 \caption{Compilation of estimated GRB luminosity distance moduli, and their errors, calibrated for redshifts $z>3.6$}
 \label{tab:grb2-1} 
\end{table*}

\section{RNN+BNN method for Cosmography} 
\label{sec:RNN-BNN}

To combine our samples at low and high redshift, in \cite{Escamilla-Rivera:2019hqt} a new computational tool was proposed for supernovae, based on machine learning (ML), called Recurrent-Bayesian Neural Networks (RNN+BNN). In this work we extend this approach to introduce our GRB sample, and apply our deep learning architecture to obtain a trained homogeneous sample comprising both SNeIa and GRBs.  The method consists of a neural network (NN) implemented with a non-linear regression process using the samples described in Sec.\ref{sec:data}.
We adopt a real target to train the NN for each data point. If the first point trained is far away from the real one, then the algorithm penalizes this point and continues the process until it reaches a \textit{true} value. When the training is carried out for the full sample, then the algorithm proceeds to minimize the loss function\footnote{Here we use a Mean Squared Error (MSE) function combined with an Adam optimizer.}.
The architecture for this NN consists of a cell where the output data, $\mu(z)$, of the previous step is used to compute the next one. Each cell is provided with the previous information of the output value $\mu$ using 
\begin{eqnarray}\label{eq:info_on_in}
    h^{<t>}&=&g(W_{h}\cdot h^{<t-1>} + W_{x}\cdot x^{<t>} +b_{a}),\\
    y^{<t>}&=&g(W_{y}\cdot h^{<t>} +b_{y}),
\end{eqnarray}
where $b$ is the bias, $g$ is the activation function, $y^{<t>}$ is the output and 
$ h^{<t>}$ and $ h^{<t-1>}$, are the hidden state and its preceding value respectively.
Moreover, RNN fails in long sequences due to the loss of previous information from the initial inputs. To improve our training performance we use a modified version of RNN as Long Short Term Memory (LSTM) cells. These cells are capable to forget or add information step by step by incrementing the number of matrix expressions for each layer as \cite{Aurelien}:
\begin{eqnarray}
    i^{<t>}&=&\sigma(W_{xi}^{T}\cdot x^{<t>}+W_{hi}^{T}\cdot h^{<t-1>}+b_{i}),\label{eq:input} \\
    f^{<t>}&=&\sigma(W_{xf}^{T}\cdot x^{<t>}+W_{hf}^{T}\cdot h^{<t-1>}+b_{f}), \label{eq:forget}\\
    o^{<t>}&=&\sigma(W_{xo}^{T}\cdot x^{<t>}+W_{ho}^{T}\cdot h^{<t-1>}+b_{o}), \label{eq:output}\\
    g^{<t>}&=&A_{f}(W_{xg}^{T}\cdot x^{<t>}+W_{hg}^{T}\cdot h^{<t-1>}+b_{g}), \label{eq:newstate}\\
    c^{<t>}&=&f^{<t>}\otimes c^{<t-1>}+i^{<t>}\otimes g^{<t>}, \label{eq:finalstate}\\
    y^{<t>}&=&h^{<t>}=o^{<t>}\otimes A_{f}(c^{<t>}) \label{eq:outputcell},
\end{eqnarray}
where $W$ are the weights of each layer, $\sigma$ is the sigmoid function that takes values between 0 and 1. Here $\otimes$ is the direct product and the superscript $T$ denotes the transpose of the quantity where it is indicated. The feed process is explicitly given in the second r.h.s term of the Eqs. (\ref{eq:input})-(\ref{eq:newstate}).

To compare different trained reconstructions of $\mu(z)$ given by the Pantheon SNeIa and GRB data, we use three different activation functions $A_f$ defined as
\begin{eqnarray}   
A_{f_{\text{Tanh}}} &=& \tanh(x), \quad \text{in} \quad (-1,1), \label{eq:tanh} \\
A_{f_{\text{ELU}}} &=&    \left\{ \begin{array}{ll}
         \alpha (e^x -1) & \mbox{for $x \leq 0$},\\
        x & \mbox{for $x > 0$}, \end{array} \right.  \text{in} \quad (-\alpha,\infty),  \label{eq:elu} \\      
 A_{f_{\text{SELU}}} &=&     \left\{ \begin{array}{ll}
         \alpha \lambda (e^x -1) & \mbox{for $x \leq 0$},\\
        x & \mbox{for $x > 0$}, \end{array} \right.  \text{in} \quad (-\alpha\lambda,\infty).  \label{eq:selu} \quad \quad
\end{eqnarray}

Furthermore, NNs have several parameters that can lead to a high probability of overfitting. To overcome this issue, we 
improve our NN by using a \textit{regularization} method  combined with \textit{Variational Dropout} (VD). These allow us to perform a regularization that can `turn off' specific neurons to avoid overfitting given certain values for the size, epochs, layers, neurons and batchsize inside the NN architecture.

\begin{figure}[h]
    \centering
    \includegraphics[width=0.62\textwidth,origin=c,angle=0]{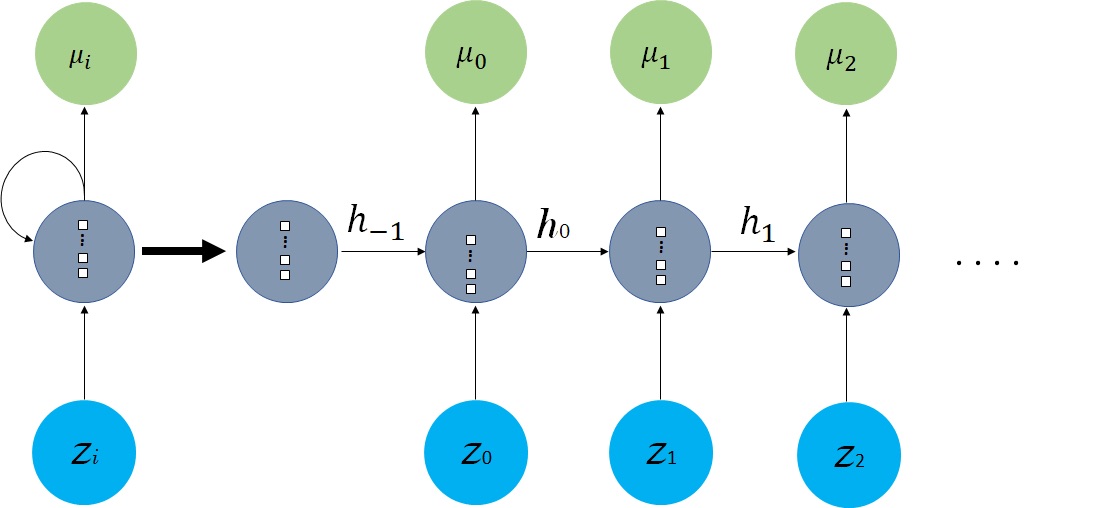}
    \caption{
Representation of our NN architecture. The input $z_i$ denotes the redshifts and the output are the distance moduli $\mu_i$ for our observable supernovae and GRBs. This NN has one layer, and the white squares inside each node denote the number of neurons $h_{i}$. In this work we take $h_{i} = 100$. As indicated by the small arrows, 
each neuron receives input information from the previous neuron. After many steps, there is some processing on the input and the computed
    result is passed to a second neuron connected through a different path, and so. Every node is associated with an activation function (e.g Tanh, ELU and SELU) which transforms the input to an output result. The output value of these functions acts as an input value to the next connected node. Final outputs can decide the class of input data. 
    }
    \label{fig:RNN_arc}
\end{figure} 

At this point, we can have some additional overfitting problems due to the transport of previous information. To solve this issue, a Bayesian Neural Network (BNN) is adapted to the NN, which can 
compute the errors on the outputs in the form of posterior distributions, making the network probabilistic.
The prior distribution on the weight function for the input point $x$ can be calculated by integrating
\begin{equation}\label{eq:prob}
    p(\mathbf{y}^{*}|\mathbf{x}^{*},\mathbf{X},\mathbf{Y})=\int p(\mathbf{y}^{*}|\mathbf{x}^{*},\mathbf{\omega})p(\mathbf{\omega}|\mathbf{X},\mathbf{Y})\mathbf{d\omega},
\end{equation}
where $p(\omega|\mathbf{X},\mathbf{Y})$ is the posterior distribution over the space of parameters and in this case $X$ and $Y$ are the redshift $z$ and the modulus distance $\mu$, respectively.

A supporting feature of our RNN+BNN network is the following: our selected  $A_{f_{\text{Tanh}}}$ is bounded in comparison to the other two activation functions. This is an ansatz that can be important in physically interpreting and representing the behavior of an observable, in our case our supernovae and GRB data. Due to the specifications of the SNeIa sample, and their use to calibrate the GRB data, we can expect that the observed differences in peak luminosities (and the six parameters for the GRBs) are closely correlated with observed differences in the shapes of their light curves.


\subsection{The method: processing the trained data} 
\label{ssec:method}

\begin{figure*}
\centering
\includegraphics[width=1.\textwidth,origin=c,angle=0]{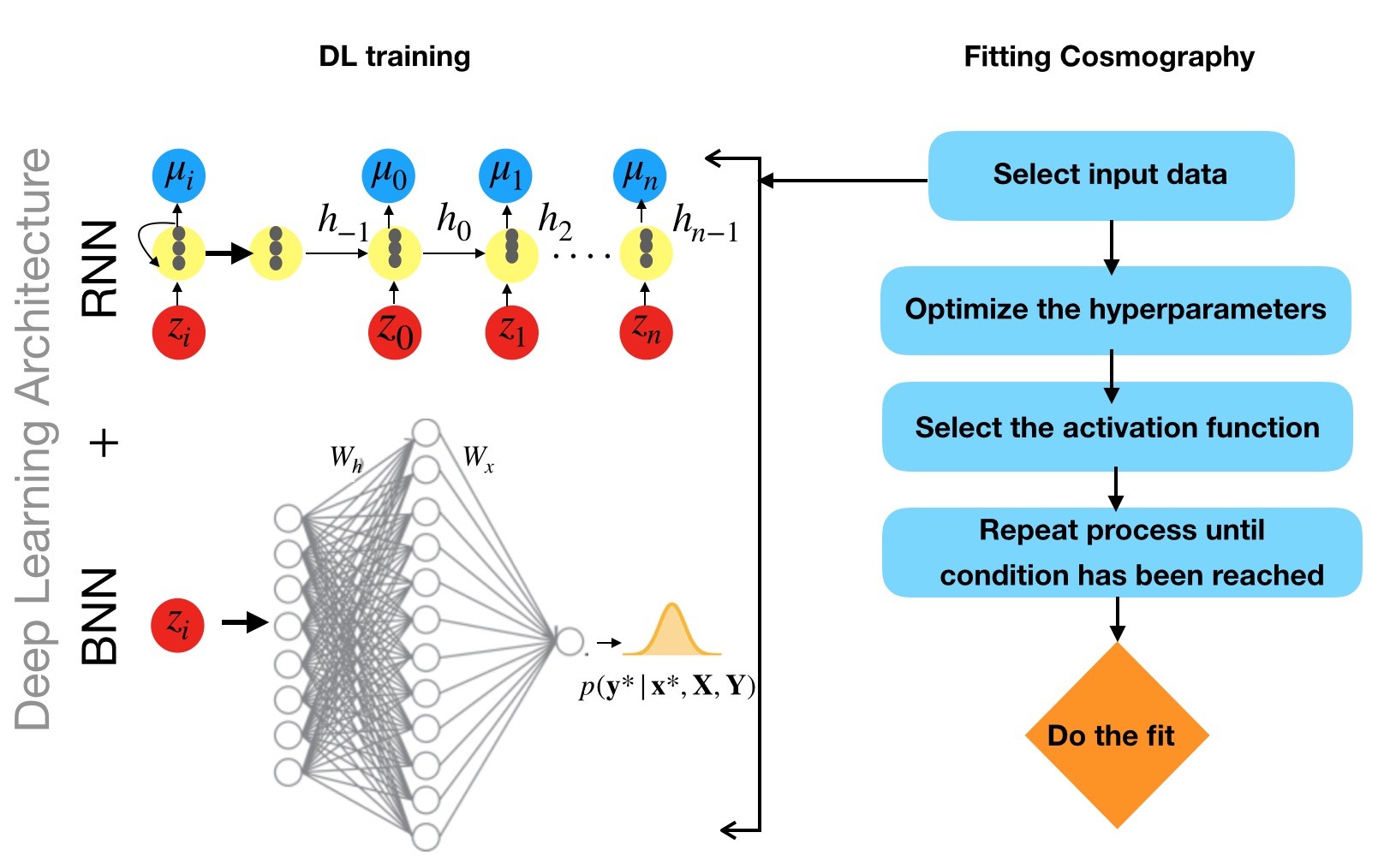}
\caption{\textit{Left Top:} Recurrent Neural Network architecture. The input $z_i$ are the redshifts and the output $\mu_i$ are the luminosity distance moduli $\mu_i$. This example figure consists of one layer and the grey circles inside each yellow node indicate the number of neurons $h_i$. The values for layers and neurons are given in (1). Each neuron receives input results from the previous neurons at each step and an activation function is associated step by step. \textit{Left Bottom:} Bayesian Neural Network architecture. This consists of one hidden layer with 12 nodes and an output layer with one node. The weight functions between input and hidden layers are given in (\ref{eq:info_on_in}), where the resulting prior distribution is given by (\ref{eq:prob}).
\textit{Right:} Schematic summarizing the steps in training the network and carrying out the fit to the data, to compute the best-fit cosmographic parameters and their credible regions.} 
\label{fig:dl_general}
\end{figure*}

\begin{figure*}
    \centering
   \includegraphics[width=7.5cm]{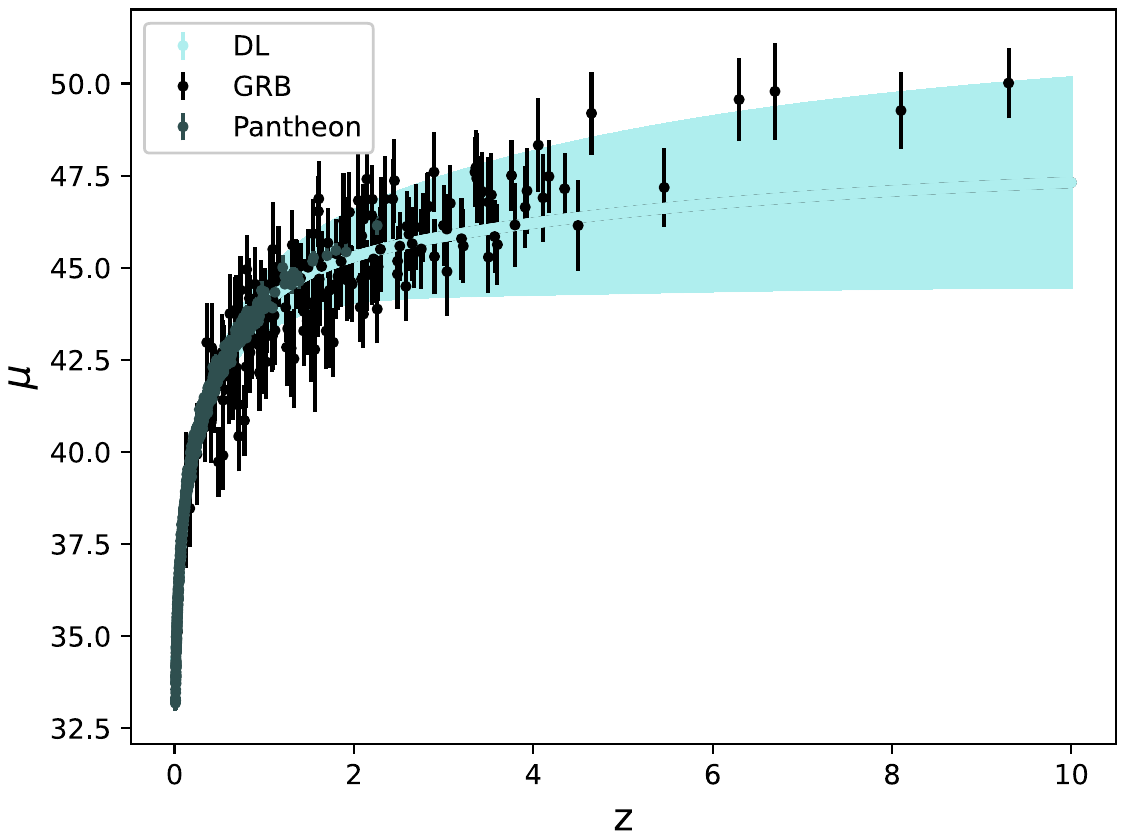}
     \includegraphics[width=7.5cm]{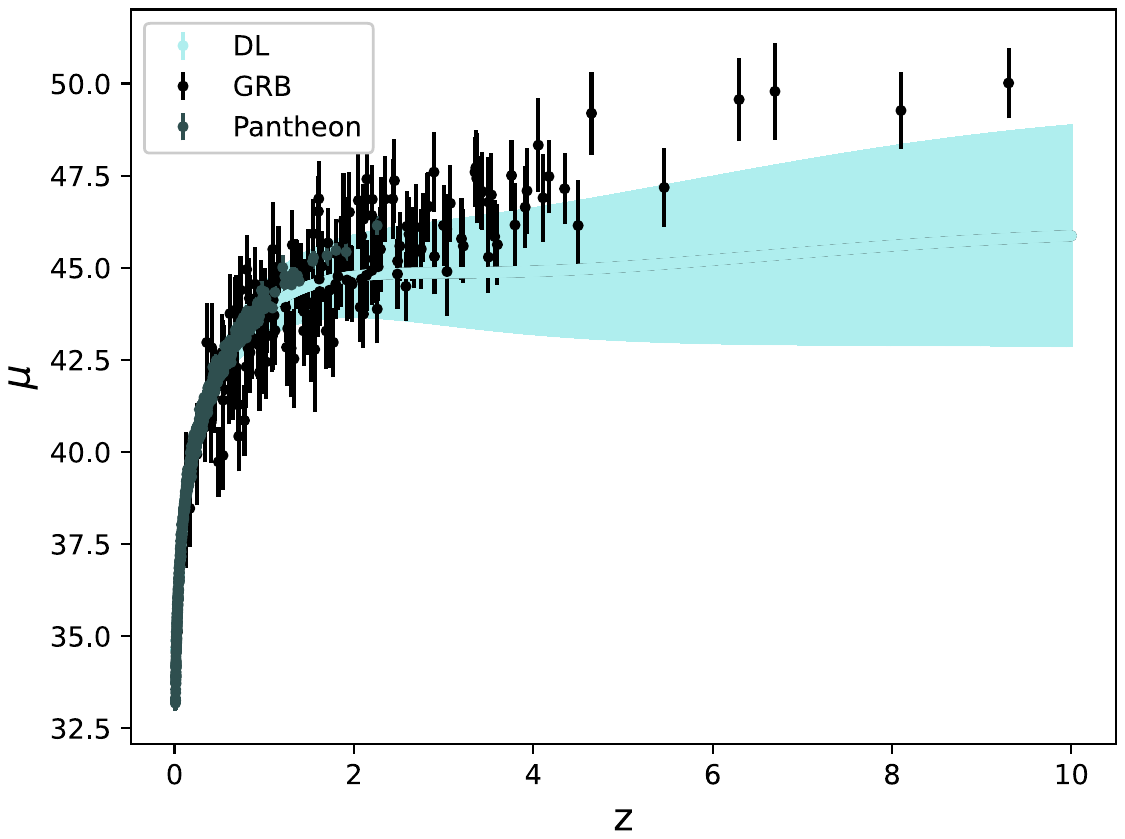}
     \includegraphics[width=7.5cm]{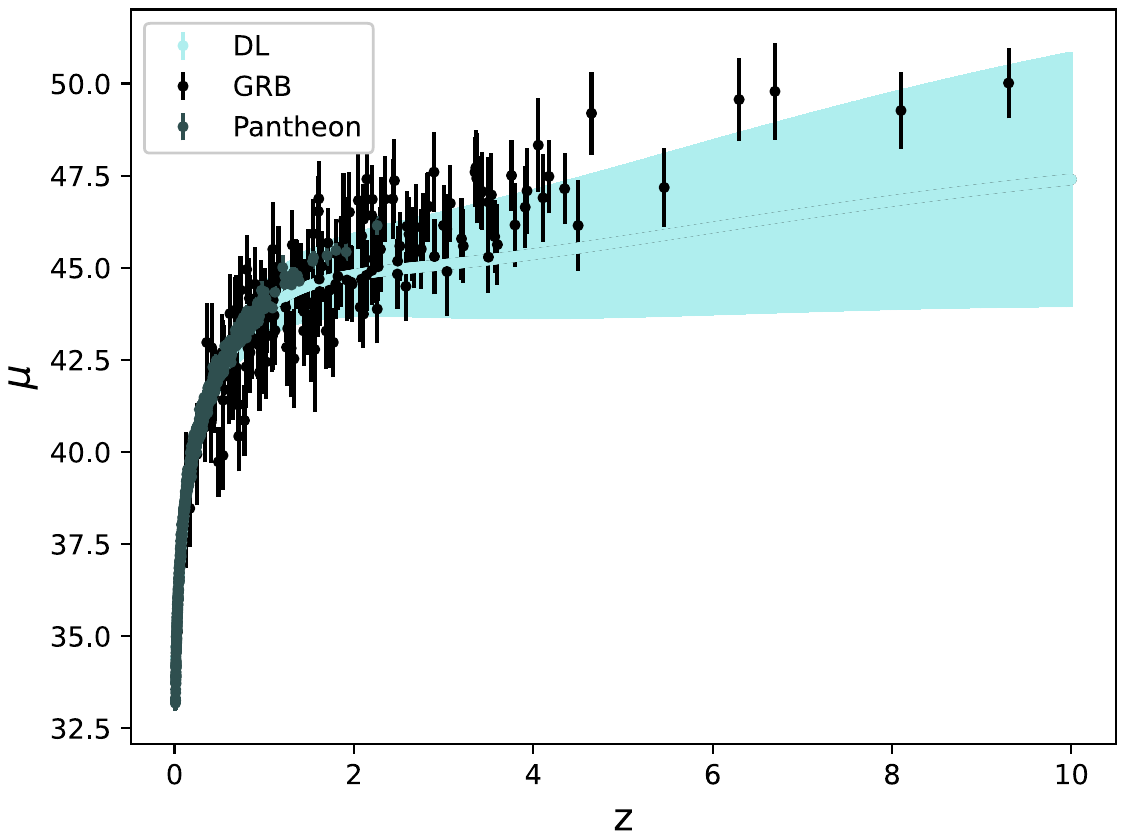}
    \caption{RNN+BNN training, up to $z = 10$, using three different activation functions: \textit{top left:} Tanh activation function; \textit{top right:} SELU activation function; \textit{bottom:} ELU activation function. 
    Black points denote the GRB data and blue points denote the Pantheon SNeIa data, while the light blue regions represent the 1-$\sigma$ confidence contours on the reconstructed $\mu-z$ relation. Notice that there is good agreement between the reconstructed relation and the high-z GRB data for the tanh activation function, but not for the other activation functions.}
    \label{fig:act_func1}
\end{figure*}

Now we describe the methodology used to train the observational data given by the Pantheon supernovae and GRBs sample using our RNN+BNN network. We analyze three different architectures that use different activation functions. The process was performed using TensorFlow\footnote{\url{https://www.tensorflow.org}}. Figure \ref{fig:RNN_arc} presents a schematic that outlines the logistics of the full method.

To perform the deep learning training we follow the methodology given in  \cite{Escamilla-Rivera:2019hqt} adapted with the Cosmography described in \cite{Munoz:2020gok}. We divide our description of this process into two sections: firstly, we describe the RNN+BNN ML architecture and secondly, we describe how we calculated the best-fit values for the cosmographic parameters that characterise the $d_L-z$ relation.

\subsubsection{
Training the Pantheon and GRBs data with RNN+BNN architecture}

    \begin{enumerate}
        \item[(1)] {\bf Design of the neural network}. We choose the three stated activation functions with the following 8 hyperparameters: 
Size = 4; Epochs = 1000; Layers = 1; Neurons = 100; Bathsize = 10; a variational dropout with an input $z$; a hidden state $h$; an output $\mu(z)$.
          \item [(2)] {\bf Organising the full sample of SNeIa + GRB data}. We ordered the data from high to low redshift, and adopted a number of steps $n=4$. The reason for this ordering is that a RN feeds itself during the training -- i.e. a network with a neuron will have a connection from the input and also from the output of its previous step, from which it feeds itself. This approach has the advantage of recording more information to train in regions where the density of data is higher (i.e. at low redshifts, for this full sample) up to regions where there is a lower density of data points (at high redshifts, for this full sample).
             \item[(3)] {\bf Computing the confidence regions via BNN}. 
            Due to the tendency of neural networks to overfit, 
             it is important to apply a form of regularisation -- i.e. to allow the algorithm to compute errors via a regularisation method. After testing several times, we found that our models could not be trained with input dropout due to the loss of information. To relax this prior,  we use an Adam optimizer.
        \item [(4)] {\bf Extending the training to higher redshifts}. After the final training, we recover the model and apply $500$ times the same dropout to our previous model. The predicted data sample with the above characteristics consists of 1209 data points in two redshift ranges: for observable ($0.01 < z < 9.3$) redshifts, and for the higher (forecast) range, $0.01 < z < 10$.  The results of our training are shown in Figure \ref{fig:act_func1}, in which we present the reconstructed $\mu(z)$ relation, using the Pantheon SNeIa and GRB data, with three different activation functions and for the extended redshift range to $z = 10$.        
        Concerning our choice of activation function, due to the physical behaviour that characterises type Ia supernovae and GRBs -- where one sees correlations between luminosity and the shapes of their light curves -- the tanh activation function of Eq.~(\ref{eq:tanh}) gave the better evolution to perform our cosmographic analysis, which we carry out in Sec.\ref{sec:results}. As expected, this better performance is seen in Figure \ref{fig:act_func1}.
          \end{enumerate}

\section{Results: SNeIa and GRB high-$z$ cosmography} 
\label{sec:results}

We now use the trained $\mu(z)$ relation, obtained following the methods described in Sec.\ref{sec:RNN-BNN}, to carry out a cosmographical analysis by using Eq.(\ref{eq:distance}) to compute the cosmographical parameters that best fit this relation. At this point we increased the number of epochs up to 1000. By doing so we are changing the weights of the network; when we increase the number of epochs to be the same as the number of times weights are changed in the NN, our analysis switches from underfitting to a better overfitting. 

\begin{figure*}
    \centering
    \includegraphics[width=0.47\textwidth,origin=c,angle=0]{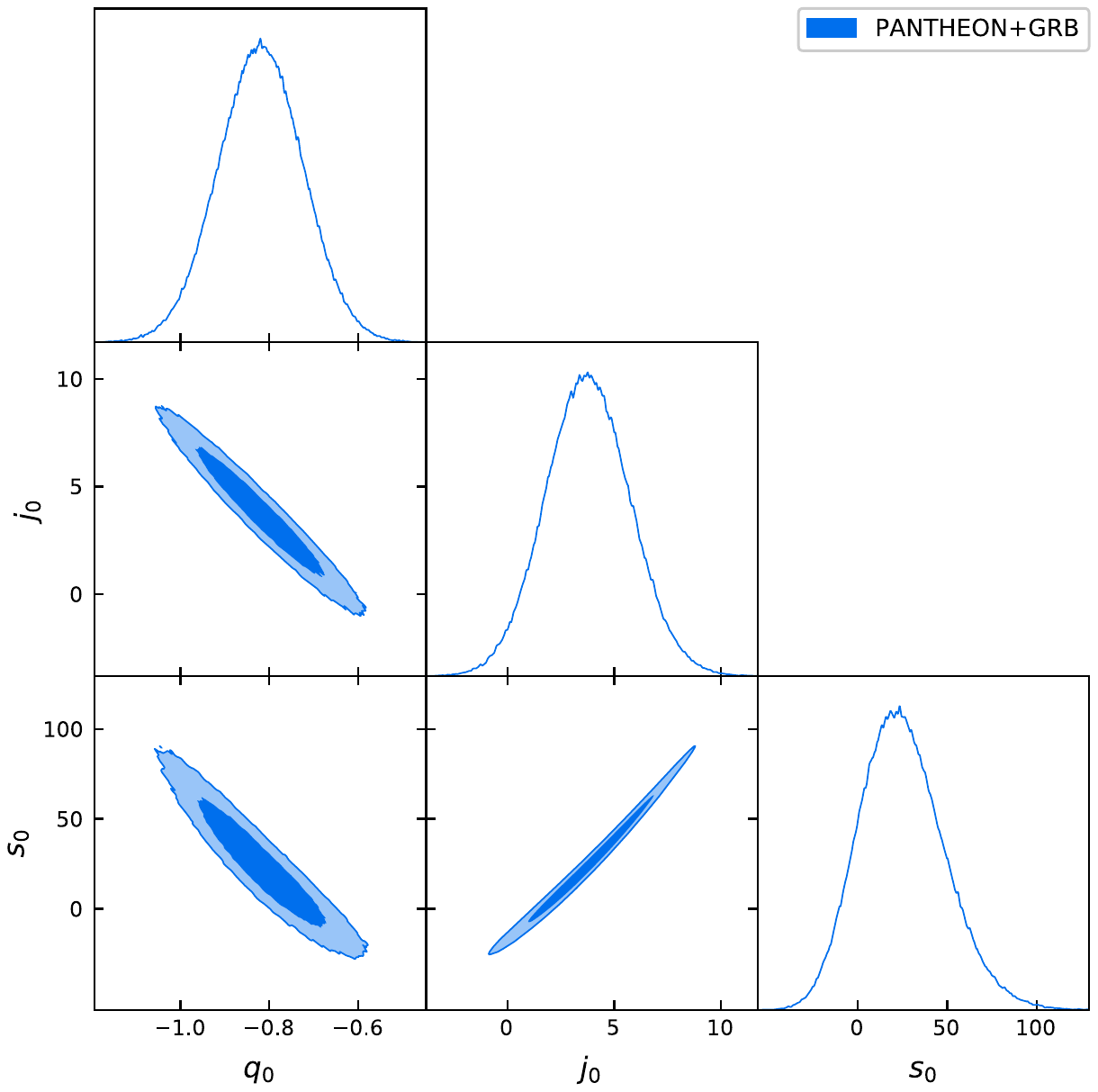}
    \includegraphics[width=0.47\textwidth,origin=c,angle=0]{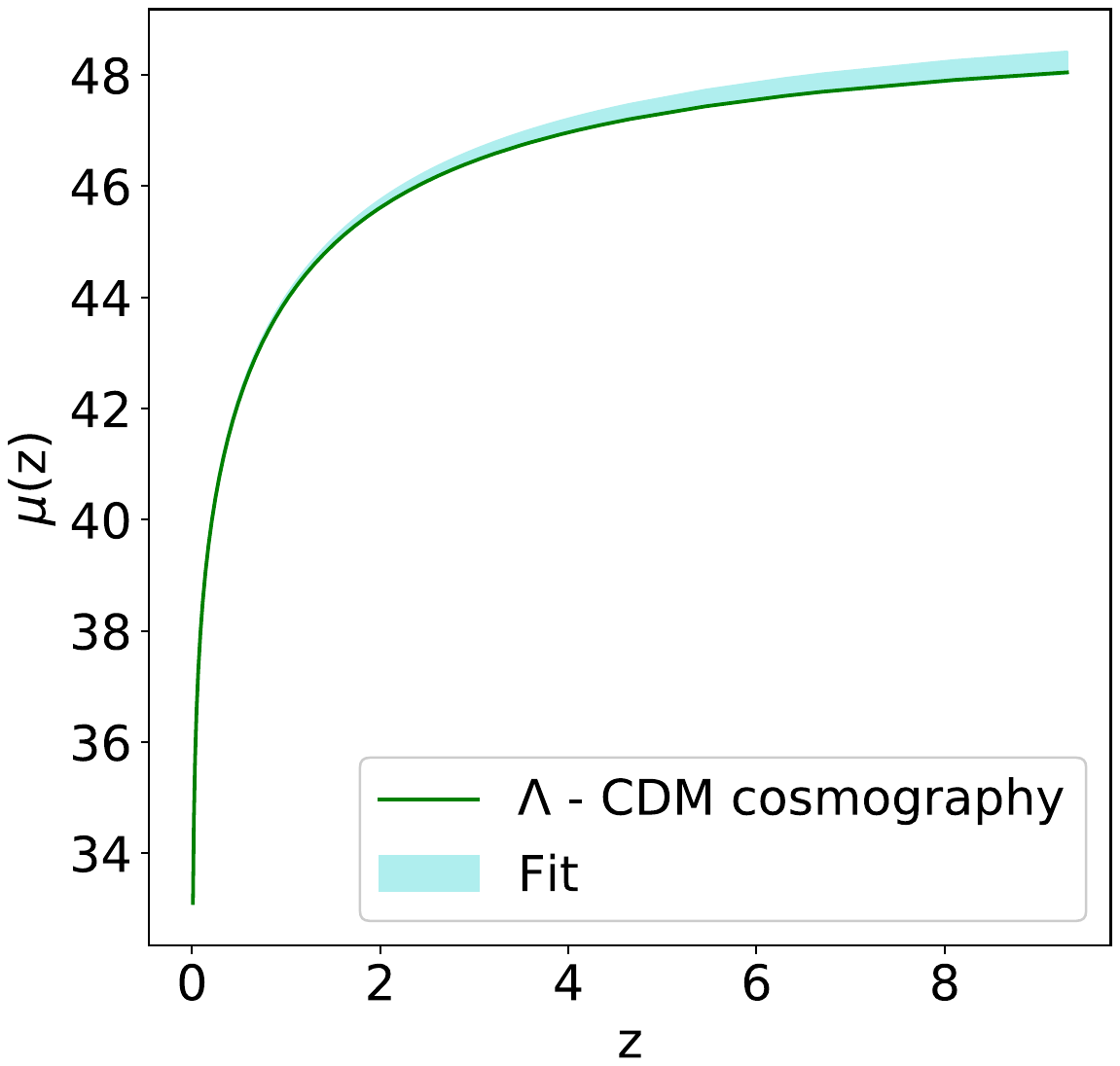}
             \includegraphics[width=0.47\textwidth,origin=c,angle=0]{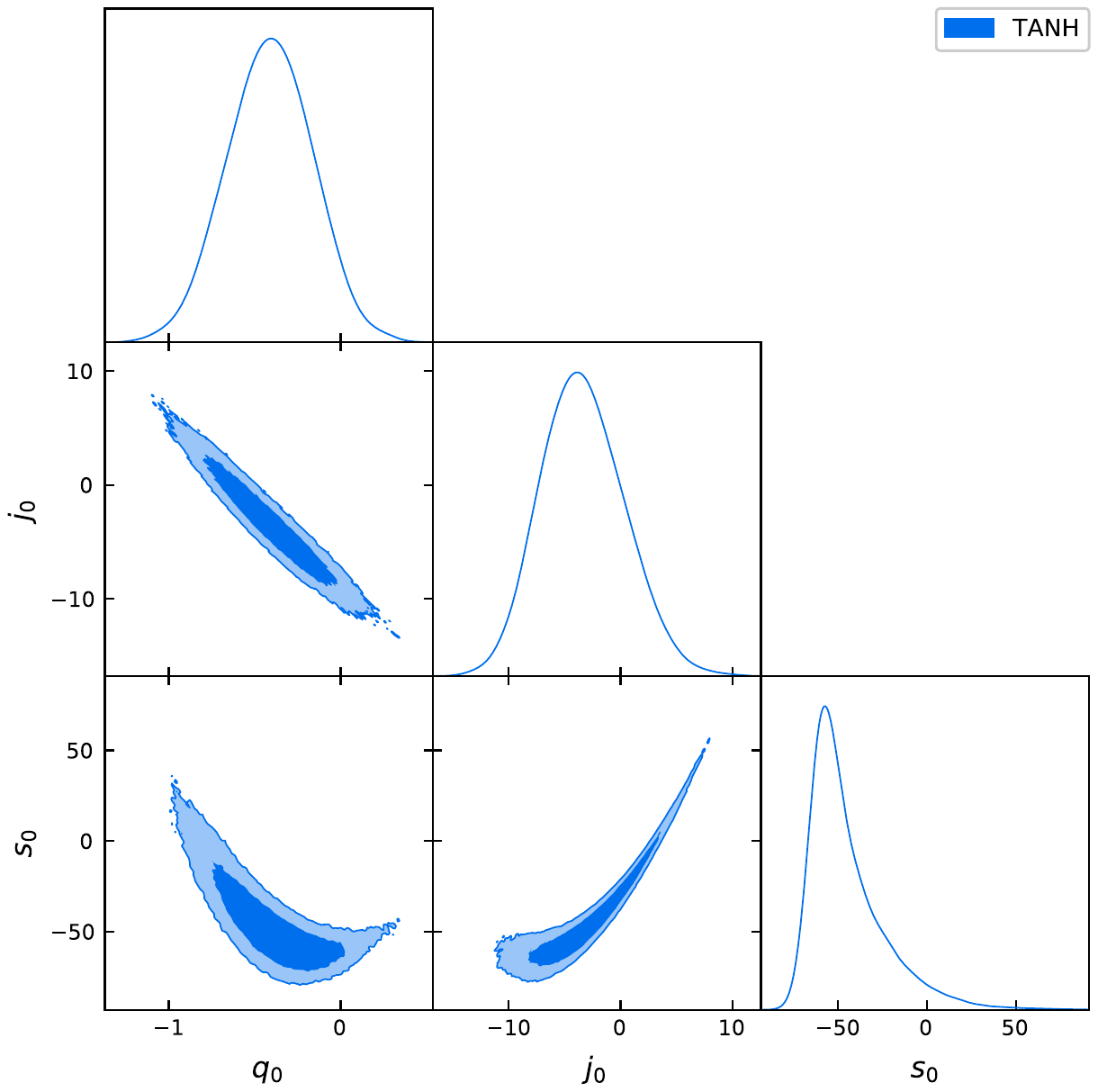}
           \includegraphics[width=0.47\textwidth,origin=c,angle=0]{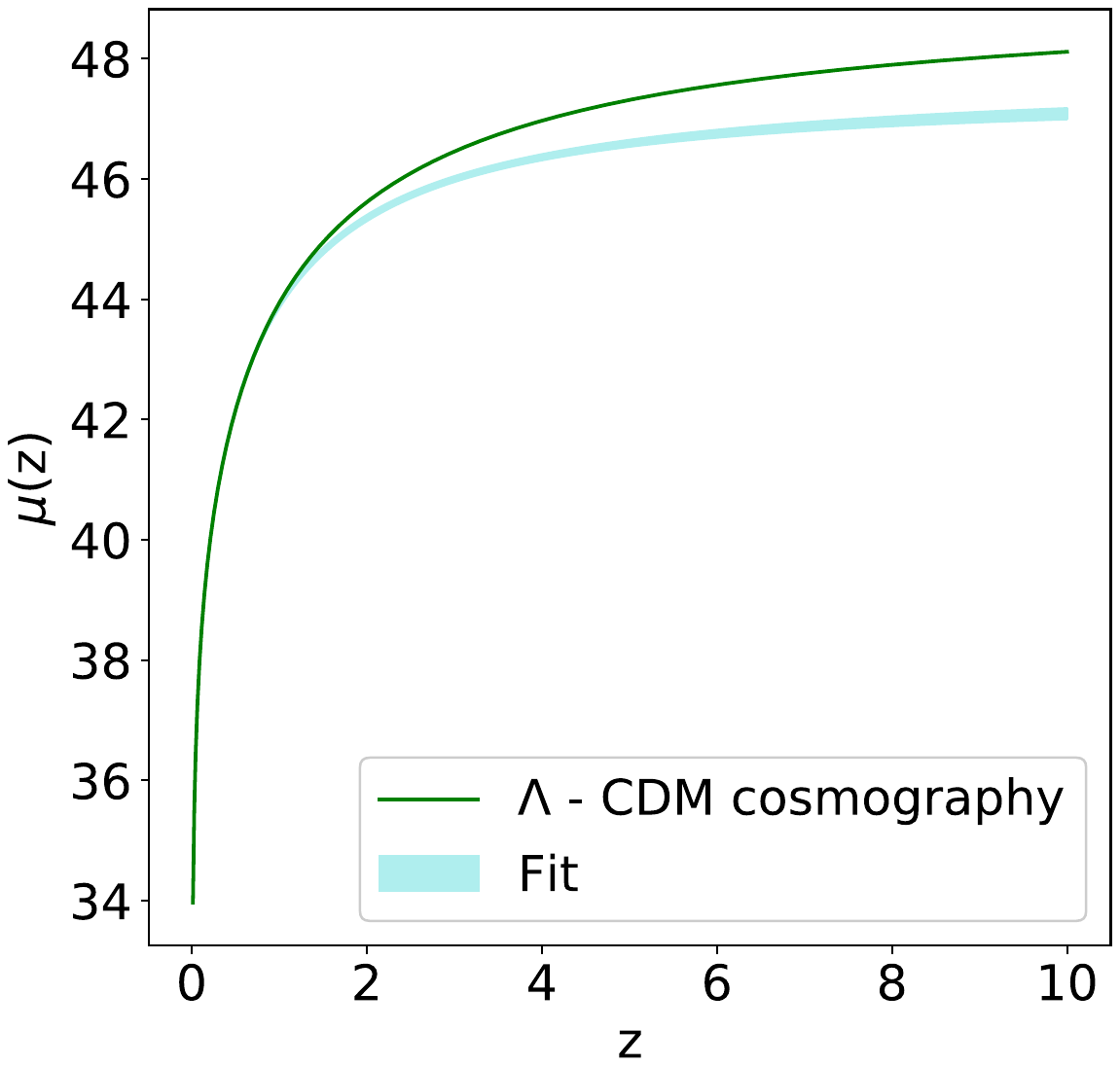}
    \caption{\textit{Left:} 2-$\sigma$ credible regions for the cosmographic parameters evaluated at the present cosmic time. \textit{Right:} Evolution of $\mu(z)$ for different cosmographic parameters. The green curves show the 
    $\mu(z)$ relations for the cosmographic parameters that correspond to the 
    $\Lambda$CDM model, while the light blue bands show the best fit $\mu(z)$ curves, and their 1-$\sigma$ errors, using the cosmographic parameter values reported in Table 4.
    The top panels show the analysis using the Pantheon SNeIa + GRB observational data, while the bottom two panels are for the analysis using Pantheon SNeIa + GRBs trained data up to $z=10$ with activation function tanh. 
    }
    \label{fig:cosmography_SNGRB}
\end{figure*}

We compute  the best-fit cosmographic parameters by modifying the publicly available codes emcee \footnote{\url{https://emcee.readthedocs.io/en/stable/}}. Corner plots showing 2-$\sigma$ credible regions for the fitted parameters are presented in the left-hand panels of Figure \ref{fig:cosmography_SNGRB}. The upper left panel shows the results obtained using the Pantheon SNeIa + GRB observational data, while the lower left panel is for the trained data up to $z = 10$ using the Tanh activation function. The numerical results of these parameter fits are also summarised in Table 4. The right hand panels of Figure \ref{fig:cosmography_SNGRB} also show the $\mu(z)$ relation, together with its 1-$\sigma$ error band, that corresponds to the best-fit cosmographical parameters derived from each data set, compared with the $\mu(z)$ relation that corresponds to the $\Lambda$CDM cosmographic model. This latter model can be derived from e.g. Planck 2018 data, for which $\Omega_m = 0.313$, and assuming flatness; this implies that $\Omega_\Lambda = 0.685$ and hence that $q_0 = -0.523$. Moreover, as noted in e.g. \cite{Capozziello:2011mcmc}, the $\Lambda$CDM model enforces $j_0=1$. Once with the values of $q_0$ and $j_0$ thus determined, we can then fix them and fit again the value of $s_0$ from our 
trained $\mu(z)$ relation to found $s_0$.

From Table 4 and Figure 4, we see that the deceleration parameter, $q_0$ is quite well constrained from the observational SNIe + GRB data (upper left panel of Figure 4) but is rather less well constrained by the trained data up to $z = 10$.  The jerk and snap parameters, $j_0$ and $s_0$ are much less well constrained by the observational data (although both are strongly correlated with $q_0$ and with each other) and the $2-\sigma$ credible regions on these parameters are again substantially larger for the trained data extended to $z = 10$ (see bottom part of Table 4). This pattern is consistent with the $\mu(z)$ graphs shown in the right-hand panels of Figure 4, where the blue and orange curves show good agreement over the full redshift range of observational data (top right panel), but the blue curve is more divergent from the orange and green curves for the trained data extended to $z = 10$ (bottom right panel).

Since we are working with a cosmographic approach via $\mu(y)$, that may result in a degeneracy in the derived cosmographic parameters, a straightforward procedure to compare two models and their parameters is the likelihood ratio test.  To compute this comparison, we can use the quantity $2\ln{\cal L}_{{\rm simple}}/{\cal L}_{{\rm complex}}$, where ${\cal 
L}$ is the maximum likelihood 
of the cosmographic model, and the subscripts `simple' and `complex' could denote, for example, predictions of $\mu(y)$ to lower and higher order respectively. The properties of the chi-squared distribution and Jeffreys' scale can then be employed to assess the significance of any increase in likelihood against the number of extra parameters (or number of data points) introduced.

As a less computationally intensive adaptation of this approach, we can also measure which models are better by taking into account how many parameters the models require to fit the data, and how good is that fit to the data, by making use of the Akaike Information Criterion (AIC) or the Bayesian Information Criterion (BIC): these criteria can be thought of as approximations to the Bayesian {\em evidence\/} for each model that are less computationally intensive to determine. Here we apply the BIC to compare our cosmographic $\mu(z)$ relation with the corresponding relation for $\Lambda$CDM.

We can compute the BIC using the relation
\begin{equation}
\mbox{BIC}= \chi^2_{\rm{min}}+d\ln N,
\end{equation}
where $N$ is the number of data points and $d$ is the number of model parameters. The quantity $\Delta\mbox{BIC}_{AB}=\mbox{BIC}_{A}-\mbox{BIC}_{B}$ can be interpreted as a measure of the evidence against model $A$ compared to model $B$. For $0\leq\Delta\mbox{BIC}_{AB}<2$ there is not enough evidence against model $A$; for $2\leq\Delta\mbox{BIC}_{AB}<6$ there is some evidence against model $A$ and for $6\leq\Delta\mbox{BIC}_{AB}<10$ there is strong evidence against model $A$.

Table \ref{tab:all-evidence} presents the results of our BIC calculations. We see that for the SNIe + GRB sample over the observed range a value of $\Delta\mbox{BIC} = 4.123$ the $\mu(z)$ relation differs markedly from that for $\Lambda$CDM.  When we consider the trained sample, extended to $z = 10$, however, we obtain a much larger value of $\Delta\mbox{BIC} > 35$, indicating much stronger evidence of a significant difference between the extended cosmographic $\mu(z)$ derived from our data and the corresponding relation for $\Lambda$CDM. This divergence, which was already apparent in Figure 4, arises because the high-redshift GRBs are still highly correlated with their low-redshift SNIe calibrators. Moreover, the divergence points towards the following robust conclusions from our analysis: 1) the BNN method is insufficient to correctly calibrate the GRB data, and 2) the RNN+BNN architecture is reliable enough to capture correctly the physical trend that dimmer SNeIa decline more rapidly after maximum brightness, and to incorporate that trend when extending our training set up to $z=10$. In order to improve its description, however, we need better quality calibration data up to $z=2.3$.

\begin{table*}
\centering
\label{tab:analysis}
\begin{tabular}{|l|c|c|c|c|c|c|c|} 
 \hline 
    \bf{Observational $z$}    
& Best-fit$\pm\sigma$                   			& $68 \%$ upper  & $68 \%$ lower  \\ \hline
   $q_0$     &   $-0.819_{-0.094}^{0.094}$             & $-0.725$  & $-0.913$     \\
    $j_0$      &  $3.754_{-1.937}^{1.977}$         &    $5.731$   & $1.817$\\
    $s_{0 }$  &  $23.576_{-22.103}^{24.706}$  &    $48.282$   & $1.473$\\
\hline 
 \end{tabular} 

 \begin{tabular}{|l|c|c|c|c|c|c|c|} 
 \hline     
\bf{Trained $z=10$}                        
& Best-fit$\pm\sigma$                   			&  $68 \%$ upper & $68 \%$ lower   \\ \hline
   $q_0$     &       $-0.410_{-0.255}^{0.249}$      &  $-0.160$     & $-0.665$ \\
    $j_0$      &    $-3.519_{-3.578}^{3.962}$    &       $0.443$   &    $-7.096$\\
    $s_{0 }$  &  $-50.287_{-12.583}^{25.340}$  &     $-24.947$   & $-62.870$ \\

\hline 
  \end{tabular} 
 
  \caption{Cosmographic best fits from (\ref{eq:modulus}) obtained  with the 1-$\sigma$ using the Pantheon SNeIa + GRBs sample and RNN+BNN trained sample. \textit{Top Table:} Results reported up to the observational redshift. \textit{Bottom Table:} Results reported up to the trained redshift.}
     \label{tab:all-best} 
\end{table*}

\begin{table*}
\centering
\begin{tabular}{|l|c|c|c|c|c|c|c|} 
 \hline 
\bf{Model (Observational)} & 					$\chi^2$                   		& BIC			& $\Delta$BIC			  \\ \hline
   $\Lambda$CDM  cosmography 			&   $1117.398   $       &    $ 1131.594  $      &  -       \\
    Cosmography (\ref{eq:modulus})      &  $1107.324  $         &     $ 1135.717   $   &  $4.123  $       \\
\hline 
 \end{tabular} 

  \begin{tabular}{|l|c|c|c|c|c|c|c|} 
 \hline 

\bf{Model (Trained $z=10$)} & 					$\chi^2$                   		& BIC			& $\Delta$BIC			  \\ \hline
   $\Lambda$CDM  cosmography   			&   $184.930     $       &    $      200.126  $      &  -       \\
    Cosmography (\ref{eq:modulus})      &  $ 12.279 $         &     $  35.080   $   &  $ 165.046$       \\
\hline 
 \end{tabular} 
  \caption{Bayesian analysis obtained using the Pantheon SNeIa + GRBs full sample. \textit{First column:} The cosmographic models in comparison to $\Lambda$CDM for our two redshift range: observational ($z:[0.01, 9.3]$) and RNN+BNN trained ($z:[0.01, 10]$). \textit{Second column:} $\chi^2$-statistics for each model. \textit{Third column:} BIC analysis. \textit{Fourth column:} $\Delta$BIC analysis. \textit{Top Table:} Results reported up to the observational redshift. \textit{Bottom Table:} Results reported up to the trained redshift.}
   \label{tab:all-evidence} 
\end{table*}


\section{Summary and Conclusions} 
\label{sec:conclusions}

In this work we have implemented a new method that uses neural networks (NNs) to calibrate Gamma-ray bursts (GRBs) detected at high-$z$ as cosmological distance indicators, combining them with a current SNeIa sample in order to trace the Hubble expansion using an approximate cosmographic relation. Using a combination of two networks -- a Recurrent Neural Network (RNN) and a Bayesian Neural Network (BNN) -- we were able to reconstruct the distance modulus-redshift relation, $\mu(z)$, of SNeIa and extend it to include the full GRB sample. Our analysis was performed over a range that extends $\mu(z)$ to high redshift, i.e. RNN+BNN: $z \in [0.01,10]$. Specifically, we simulated two types of data with the BNN: first using the same redshift range as the observational data and second when extending the training up to $z=10$.

This cosmographic approach required a very minimal set of assumptions that does not rely on the dynamical equations for a specific theory of gravity. Our NN analysis considered three different activation functions to train our models: ELU, SELU, and Tanh.  We found that the trend obtained by the SELU function (\ref{eq:selu}) does not follow the physical behaviour with the full sample of SNeIa + GRB at higher redshift; therefore we discarded this activation function. By comparing the ELU function (\ref{eq:elu}) with the Tanh function (\ref{eq:tanh}), we obtained a significant $\chi^{2}$-difference in the performance of these two activation functions -- with the Tanh function clearly favoured. Due to the random initialization of weights in the RNN, we employed a cross-validation method to find the model with the lowest $\chi^{2}$ for the Tanh function. 

From the results of our analysis, as summarised in Figure \ref{fig:cosmography_SNGRB}, we can draw the following conclusions:
\begin{itemize}
\item For the trained SNeIa + GRB using RNN+BNN only up to the observed redshifts reported in the literature (i.e. for Pantheon up to $z=2.3$ and for the GRB sample up to $z=9.3$), the best-fit cosmographic relation obtained with these data indicates that the training is reliable up to $z=2.3$. Specifically, the best-fit $\mu(z)$ is essentially indistinguishable from the $\Lambda$CDM cosmographic relation up to $z=2.3$ and deviates at only about 1-$\sigma$ from the latter relation at higher redshift. We note that for $z<2.3$ the density of data points is larger compared with that for the observable points at higher redshifts. Hence, for $z > 2.3$ the slight deviation of our best-fit $\mu(z)$ relation from the cosmographic relation derived from the $\Lambda$CDM parameters is likely the result of two factors: 1) the lower density of data points and the larger errors on the estimated luminosity distance moduli of GRBs at higher redshift, and 2) the possibility that the cosmographic relation derived from the $\Lambda$CDM model parameters is inadequate to explain the observable GRB data for $z>2.3$.
 
\item For the trained SNeIa + GRB, using RNN+BNN to extend the training up to $z=10$, again we see that the training is fully reliable for $z<2.3$. At higher redshifts, however, the reconstructed $\mu(z)$ starts to deviate more strongly from the cosmographic relation derived from the $\Lambda$CDM model than was the case for our analysis using only the observable redshift range. This larger deviation may be due in part to our choice of activation function, which sets a bias on the training itself. Thus, while the trained data over the observed redshift range does provide support for $\Lambda$CDM as the preferred model, this result may be biased due to the low density of data points above $z>3.6$ (or even the lack of data between $z=7$ and $z=8$, see Table \ref{tab:grb2-1}), and we can expect that bias to be more important when the RNN+BNN training is extended up to $z=10$.  Notwithstanding this, of course the best-fit $\mu(z)$ relation for $z>3.6$ may also be telling us that the standard $\Lambda$CDM cosmography {\em is\/} inadequate at high redshift.
\end{itemize}

The work presented here represents a powerful, new NN methodology for combining and calibrating SNeIa and GRB datasets, but also emphasises the requirement of more \textit{real} data at higher redshifts that can help the NN to identify the physics trends underlying the cosmographic relations traced out by these datasets. With such additional calibration data, we believe that this new NN approach can successfully calibrate GRB distance information over an extended redshift range, up to $z=10$, and thus can help to compare the cosmography inferred from supernovae observations with the cosmography inferred at higher redshift. 


\acknowledgments

CE-R acknowledges the \textit{Royal Astronomical Society} as FRAS 10147 and supported by DGAPA-PAPIIT-UNAM Project TA100122. 
CDZM was supported by DGAPA-PAPIIT-UNAM Project IA100220.
M. H. is supported by the Science and Technology Facilities Council (Ref. ST/L000946/1). 
The Authors thank the anonymous Referee for the constructive critics on this paper.



\end{document}